# Compelling experimental evidence of a Dirac cone in the electronic structure of a 2D Silicon layer

Sana Sadeddine[1], Hanna Enriquez[1], Azzedine Bendounan[2,*], Pranab Kumar Das[3,4], Ivana Vobornik[3], Abdelkader Kara[5], Andrew J. Mayne[1], Fausto Sirotti[2], Gérald Dujardin[1], and Hamid Oughaddou[1,6,*]

[1]Institut des Sciences Moléculaires d'Orsay, ISMO-CNRS, Bât. 210, Université Paris-Sud, F-91405 Orsay, France

[2]TEMPO Beamline, Synchrotron Soleil, L'Orme des Merisiers Saint-Aubin, B.P.48, 91192 Gif-sur-Yvette Cedex, France

[3]CNR-IOM, TASC Laboratory, AREA Science Park Basovizza, I-34149 Trieste, Italy

[4] International Center for Theoretical Physics (ICTP), Strada Costiera 11, I-34100 Trieste, Italy

[5]Department of Physics, University of Central Florida, Orlando, FL 32816, USA

[6]Département de physique, Université de Cergy-Pontoise, F-95031 Cergy-Pontoise Cedex, France

**ABSTRACT:**

**The remarkable properties of graphene stem from its two-dimensional (2D) structure, with a linear dispersion of the electronic states at the corners of the Brillouin zone (BZ) forming a Dirac cone. Since then, other 2D materials have been suggested based on boron, silicon, germanium, phosphorus, tin, and metal di-chalcogenides. Here, we present an experimental investigation of a single silicon layer on Au(111) using low energy electron diffraction (LEED), high resolution angle-resolved photoemission spectroscopy (HR-ARPES), and scanning tunneling microscopy (STM). The HR-ARPES data show compelling evidence that the silicon based 2D overlayer is responsible for the observed linear dispersed feature in the valence band, with a Fermi velocity of $v_F \sim 10^{+6}\ m.s^{-1}$ comparable to that of graphene. The STM images show extended and homogeneous domains, offering a viable route to the fabrication of silicene-based opto-electronic devices.**

The discovery of graphene[1-3] opened the way for exploring new 2D materials made from boron[4], silicon[5-9], germanium [6,10-12], phosphorus[13], tin[14], and transition metal di-chalcogenides[15,16]. However, theory predicts that along with graphene[1-2,17], only graphene-like 2D structures of silicon, germanium (silicene and germanene)[6], boron[18], and $MoS_2$[19] should have a Dirac cone. Experimentally, only graphene has unambiguously shown a Dirac cone[1-2]. Silicon has attracted much attention over the last few years[8,9]. The critical discussion has focused on whether graphene-like silicon (silicene) exists, and what its electronic properties are. Earlier theoretical studies predicted a puckered honeycomb structure for free-standing silicene with electronic properties resembling those of graphene[6].

Experimentally, silicene has been obtained by epitaxial growth on crystalline surfaces[5,7-9]. The majority of studies were performed on Ag surfaces, where the single silicon layer forms an assembled parallel array of one-dimensional (1D) nano-ribbons (NRs) on Ag(110)[8,9,20], and a highly ordered silicene sheet on Ag(111)[7-9]. Very recently the fabrication of silicene-based field effect transistors (FETs) operating at room temperature (RT), was reported[21]. Since silicene has been obtained only by growth on supporting metallic substrates, the obvious question arises as to whether it preserves the Dirac cone or not. These are the key questions: can we separate the electronic structure of silicon from that of the underlying substrate, and can we observe the Dirac cone of silicene?

Reports of the existence of Dirac fermions in silicene on the Ag(111) surface using angular-resolved photoemission spectroscopy (ARPES) and scanning tunneling spectroscopy (STS) have been published[22,23]. However, several recent studies point to the absence of the Dirac cone[24-26], as a result of the strong silicon-silver interaction[27,28], which alter the intrinsic electronic properties of silicene[24-26]. These studies show that the linear dispersion observed previously stems from the electronic structure of silver. The absence of a Dirac cone in silicene on silver has challenged research groups to explore other potential substrates having weaker interactions with silicene.

The Si-Au system prefers phase separation over alloy formation[29] since the silicon surface energy of 1.200 J/m$^2$ is smaller than that of Au at 1.506 J/m$^2$. In addition, there is a good lattice match since the four-nearest neighbor Au-Au distance (1.156 nm) coincides with three-unit cells of the (111) surface of silicon (1.152 nm). These conditions suggest that the formation of a silicene sheet is favorable on the Au(111) surface.

Here, we present a comprehensive experimental study of the growth of silicon on an Au(111) surface. The experiments reveal the formation of an extended 2D silicon sheet with long range order using low energy electron diffraction (LEED) and high resolution angle resolved photo-emission spectroscopy (HR-ARPES). The HR-ARPES measurements display unambiguously the presence of the Dirac cone at the K-point with an estimated band gap of about 0.5 eV. X-ray photoemission spectroscopy (XPS) measurements reveal that the silicon has a single chemical environment, while the scanning tunneling microscopy (STM) images show a highly ordered layer. These results are the first clear evidence for the formation of silicene presenting a Dirac cone.

Figure 1a displays a LEED pattern of the bare Au(111) surface. The signature of the herringbone structure is shown by the satellite spots visible around the substrate spots. Figure 1b displays the LEED pattern obtained after the deposition of 1 silicon ML on the Au(111) surface (held at 260°C, see Methods for details). Some diffraction spots are located at the nodes of a 12x12 reconstruction as indicated by the hexagonal grid. The spots belonging to this "by-12" along the high symmetry directions are highlighted by the red arrows. All other spots can be assigned to the diffraction of an incommensurate rectangular cell (0.73 nm x 0.92 nm) rotated by 19° compared to the principal Au(111) direction, that forms twelve domains due to the six-fold rotation and two mirror symmetries of the substrate. The agreement between the observed spots given by the (0.73 nm x 0.92 nm) rotated by 19° in Figure 1b and the corresponding simulation shown in Figure 1c is remarkably good for such a complex pattern. Note that the simulated pattern *does not* show the 12x12 spots indicated by red arrows in Figure 1b, as this superstructure was not included in Figure 1c for clarity.

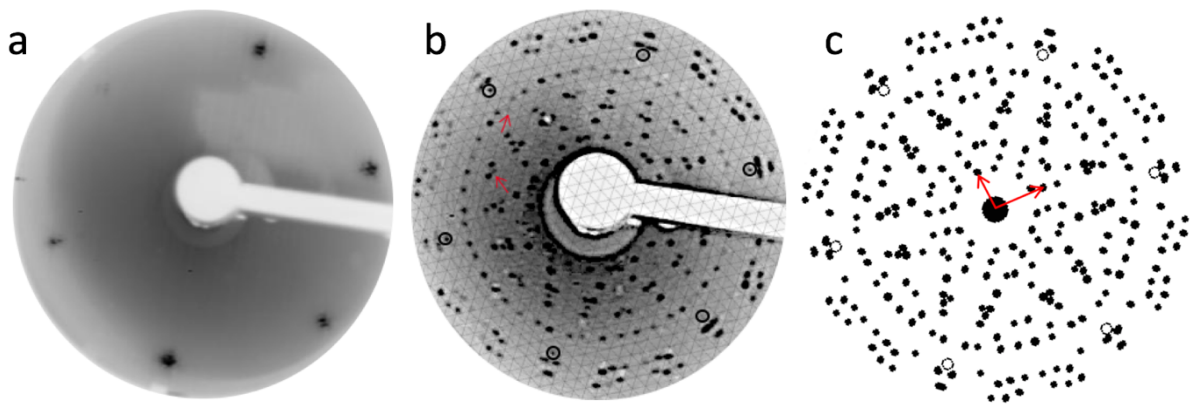

Figure 1: LEED patterns recorded at Ep = 50eV corresponding to the bare Au(111) (a) and to 1 Si ML deposited on Au(111) kept at 260°C (b). The signature of the herringbone structure is shown by the small satellite spots visible around the integer order spots on (a). The LEED pattern shown on (b) corresponds to the coexistence of two superstructures: a 12x12 reconstruction (shown by hexagonal grid) and an incommensurate rectangular cell (0.73 nm x 0.92 nm) rotated by 19° compared to the principal Au(111) direction presenting twelve domains (c). A simulated LEED pattern corresponding to a (0.73 nm x 0.92 nm) rotated by 19°, including the twelve domains. The reciprocal unit cell is drawn with red arrows and the 1x1 spots of the substrate with black hollow circles. The red arrows on (b) show spots located along the high symmetry direction of Au(111) belonging to the 12x12 structure.

ARPES measurements were made on 1 ML of silicon and were recorded at low temperature (78 K) at two photon energies, 147 eV and 66 eV, along the $\overline{\Gamma}$-$\overline{K}$ direction of the substrate. On the bare Au(111) we observe the Au *sp* band as indicated in Figure 2a and 2c. However, after deposition of 1 Si ML, a clear conical dispersion of Si bands is observed at $k_{\parallel}$ = 1.33 Å$^{-1}$ at both photon energies (Figures 2b and 2d). The Au *sp* branch is still observed at the photon energy of 147 eV together with the silicene Dirac cone feature, whereas it is absent at 66 eV due simply to a cross section effect. The conical dispersion is unambiguous evidence for the existence of Dirac fermions in the silicon ad-layer indicating that the silicon atoms have *sp²* hybridization and hence have silicene characteristics. A single *sp²* band is observed without any spin-orbit splitting, although this latter was expected in recent theoretical studies[30,31]. The failure to detect such a splitting is probably due to the limited angle and energy ARPES resolutions.

Now, the expected Dirac cone of the 1x1 silicene should be located at $\frac{9}{12} \times \overline{K}_{Au}$ = 1.09 Å$^{-1}$ (with $\overline{K}_{Au}$ = 1.45 Å$^{-1}$) based on the match between the four nearest neighbor Au-Au distances (1.156 nm) and the three unit cells of the (111) surface of silicon (1.152 nm). However, in our case it is located at $k_{\parallel}$ = 1.33 Å$^{-1}$ (Figure 2e) which corresponds exactly to $\frac{11}{12} \times \overline{K}_{Au}$ suggesting the existence of a "by-12" superstructure of silicene compared to the Au(111), in good agreement with the LEED measurements. The appearance of the Dirac cone at $k_{\parallel}$ = 1.33 Å$^{-1}$ can be the result a band-folding induced by a "by-12" superstructure. Figure 2e shows the different BZ of Au, the 1x1 silicene, and the 12x12 structure of silicene. The existence of the linear dispersion at all six equivalent $\frac{11}{12} \times \overline{K}_{Au}$ has been checked and confirmed. A honeycomb structure is known to produce a photoemission interference effect [32], so that one branch appears brighter than the other. This effect has been already observed for graphene[32].

In Figure 2b, both the Au *sp* band and the Dirac cone are observed simultaneously in contrast to the case of the (4x4) silicene on Ag(111)[22]. This indicates that the

silicene-Au(111) interaction is not as strong as for silicene on Ag(111)[25]. The linear dispersion can be described by E = $\hbar.v_F.k$, where $v_F$ is the Fermi velocity. From the linear dispersion in Figure 2 we obtain a Fermi velocity for the silicene layer of $v_F = 1.3 \times 10^{+6}\ ms^{-1}$, comparable to the one found for graphene[32].

The apex of the cone is located at 0.5 eV below the Fermi level. Since the $\pi^*$ cone could not be detected, we can assume that the $\pi$ - $\pi^*$ gap opening is at least equal to 0.5 eV. Freestanding silicene is expected to have a zero gap[6]. However, the opening of the band gap could result from a weak interaction with the Au(111) substrate such as a small lattice mismatch or the existence of a surface dipole, which would not affect the intrinsic electronic states of silicene. Such effect has been already observed for graphene[3,33-35].

The backfolding of the sp bands of Au (3D) can be ruled out because they would not show any opening of a gap. Therefore, the presence of a gap here leading to a conical dispersion with a "Λ shape" can be attributed to the 2D character of the band induced by the deposition of one silicon monolayer.

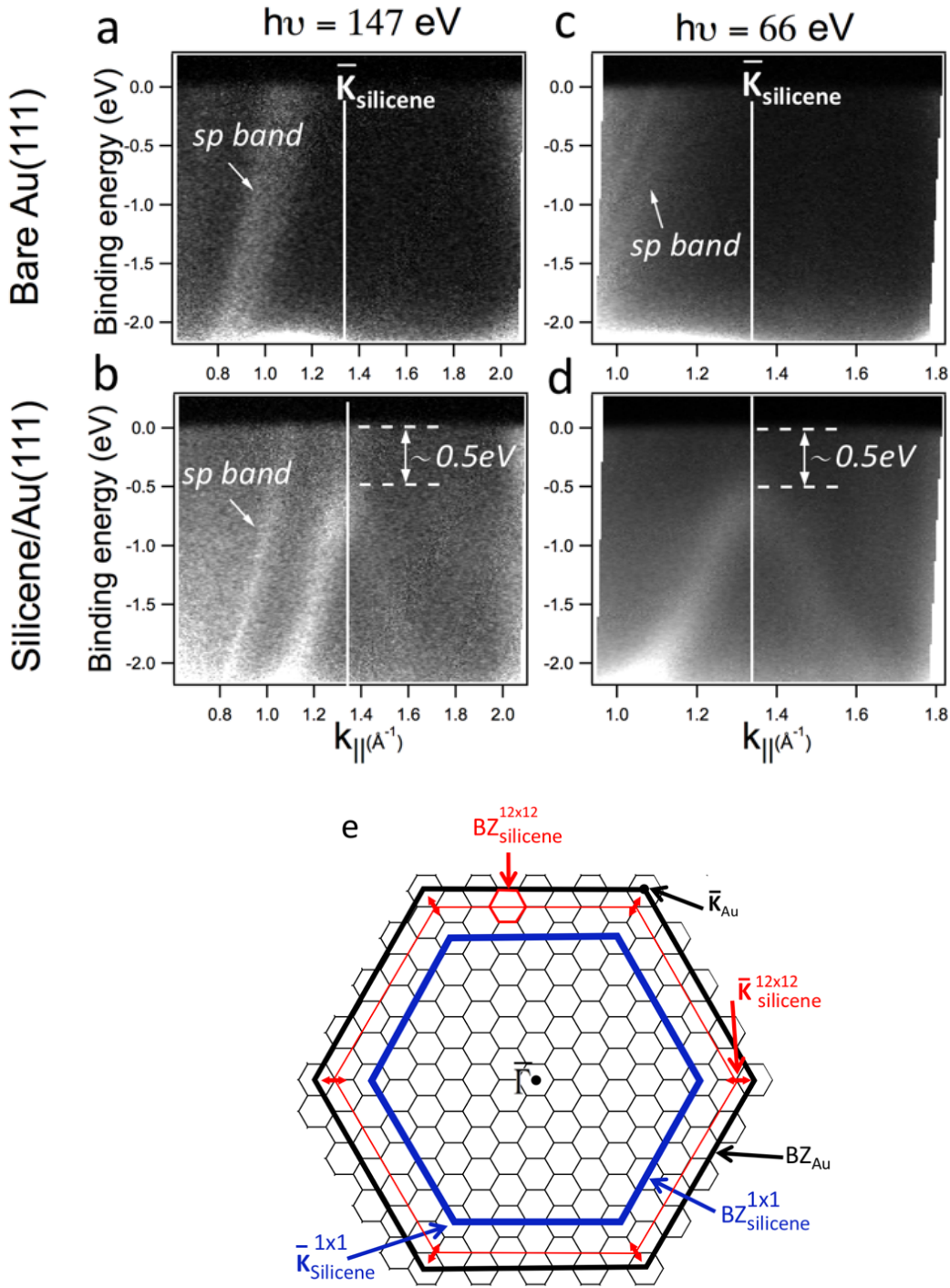

Figure 2: High resolution ARPES measurements on clean Au(111) (a,c) and upon deposition of one silicene monolayer (b,d). The experiments were performed at 78 K using two photon energies (66 and 147 eV). On silicene the presence of Dirac-like feature at the K-point of the silicene SBZ is observed clearly, with a band gap of 0.5 eV. (e) Schematic representation of Au and silicene SBZ, in which the principal high symmetry points are indicated. At a photon energy of 147 eV, the Au sp branch is clearly observed on both bare Au(111) and silicon monolayer covering Au(111) surfaces. However, at 66 eV the same branch occurs very weak on Au(111) and disappears completely after deposition of the silicon monolayer. This difference is due to a cross section effect.

The Si 2p core level spectra of the adsorbed silicene sheet, recorded at bulk (0° normal emission) and surface sensitive (75° off-normal emission) geometries, are shown in Figure 3. In both cases the Si 2p spectra are reproduced with one spin–orbit split component located at 99.825 eV indicating that all silicon atoms have only one chemical environment. The fact that both spectra are very similar also indicates that all silicon atoms are nearly at the same height. This result supports the interpretation that the interaction between silicene and Au surface is weak ruling out the formation of any alloy.

The best fit was obtained with a 180 meV Gaussian profile and a 70 meV Lorentzian profile, while the spin-orbit splitting is 0.605 eV, and the branching ratio is 0.52. It was necessary to include an asymmetry parameter of 0.046. This indicates the Silicon-Gold system has a metallic character. The very narrow total width of 280 meV indicates a high ordering of the Si atoms within the silicene sheet.

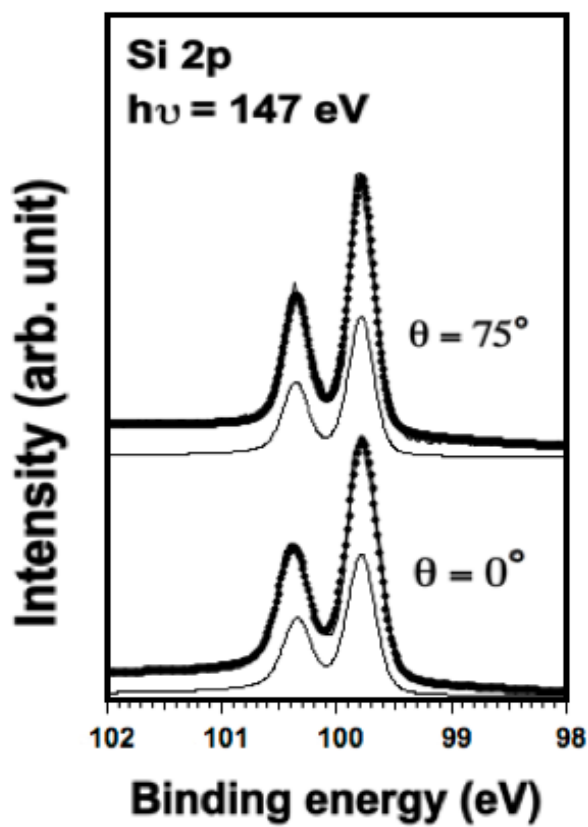

Figure 3: Si 2p core levels spectra (dots) and their de-convolutions (solid line overlapping the data points) with one asymmetric component recorded after deposition of 1 Si ML on Au(111) at hv = 147eV. 0° normal emission and 75° off-normal emission correspond to bulk and surface sensitive geometry, respectively. The final fitting parameters used are: 150 meV Gaussian width, 80 eV Lorentzian width, a spin-orbit splitting of 0.605 eV, a branching ratio of 0.52, and an asymmetry parameter of 0.046.

The STM data were recorded at low temperature (78 K). The well-known STM image of the "herringbone" reconstruction of the clean Au(111)[36] is shown in Figure 4a. The STM topography in Figure 4b shows a part of the "herringbone" reconstruction at atomic resolution. The STM topography in Figure 4c shows the Si monolayer which covers the surface terraces forming a Moiré structure. The Moiré is aligned along the "herringbone" lines (parallel to the $[11\bar{2}]$ direction) where all the underlying Au atoms occupy the same close-packed structure with a periodicity of approximately 5.4 nm. However, perpendicular to the $[11\bar{2}]$ direction, the Moiré is not perfectly aligned (see broken lines) most probably because the underlying gold atoms occupy either *fcc* or *hcp* sites. This suggests that the "herringbone" reconstruction of the bare Au(111) is still present beneath the silicon layer. Figure 4d displays a zoom of the STM image of Figure 4c. This image shows a trigonal atomic pattern with a rectangular unit cell of 0.73 nm x 0.92 nm rotated by 19° with respect to Au $[01\bar{1}]$ direction in excellent agreement with the LEED measurements. In addition, the "by-12" observed in the LEED pattern and in ARPES measurements is not visible in the STM images probably due to a weak electronic coupling between the silicene sheet and the Au substrate.

The two protrusions within the unit cell in the STM image are different because they are rotated with respect to each other as highlighted by the ellipses (see Figure 4d). Note that each protrusion does not correspond to a single atom. This trigonal atomic structure is compatible with a hexagonal silicon lattice. In many other cases, trigonal lattices derived from a hexagonal lattice have been observed (bilayer graphene[3], Si/Ir[37], Ge/Al[10], etc.). Even though, the precise atomic structure is beyond the scope of this paper and is still under investigation, these STM topographies confirm that the Au reconstruction is still present and that the 2D silicon layer forms a new trigonal atomic structure compatible with the LEED measurements.

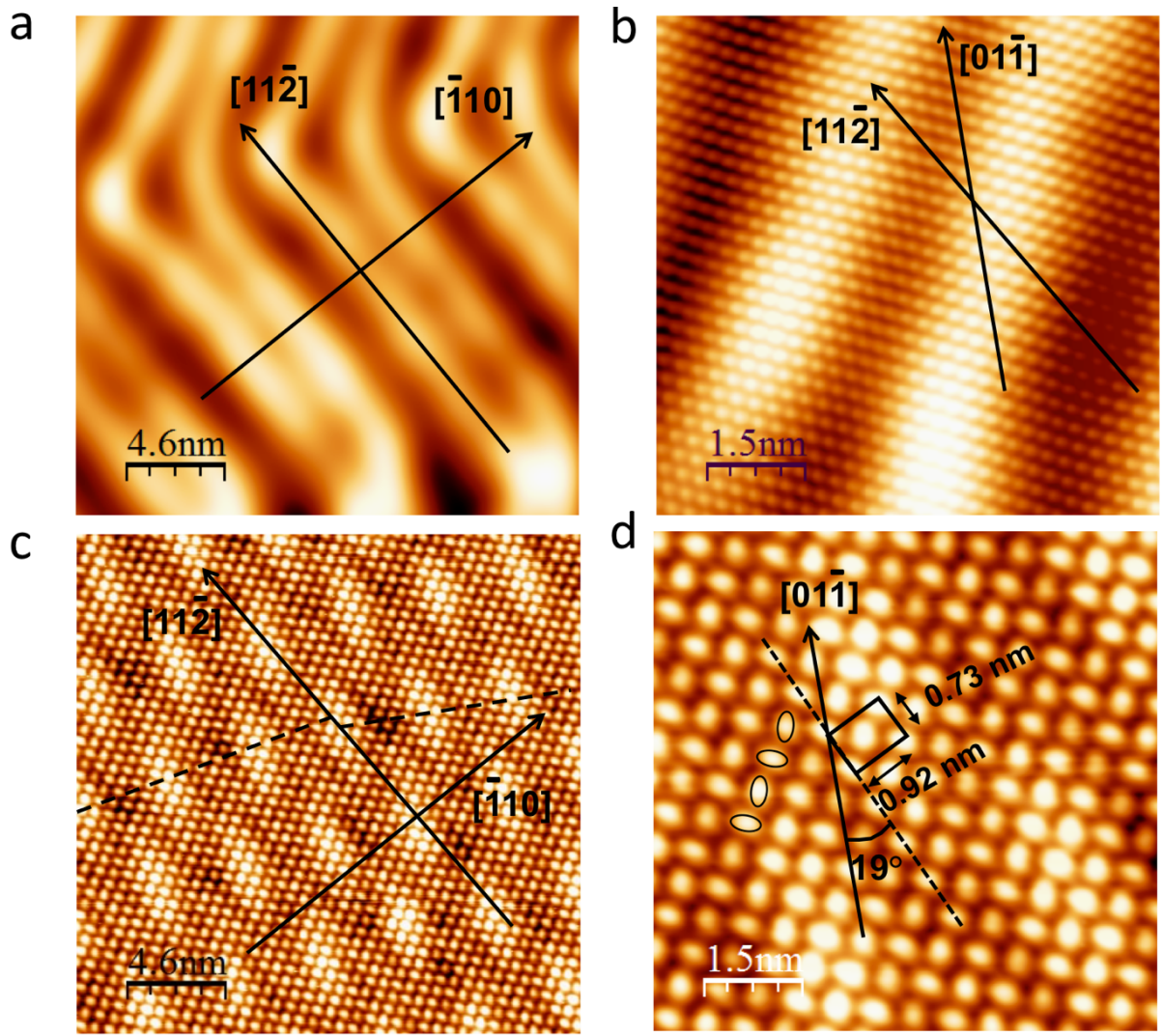

Figure 4: The STM images corresponding to (a) the bare Au(111) showing "herringbone" reconstruction (U = -0.34 V, I= 1 nA), (b) the bare Au(111) showing a part of "herringbone" reconstruction at atomic resolution (U = -0.34 V, I= 3.2 nA), (c) 1 ML of silicon deposited on Au(111) (U = -1V, I= 1 nA). The moiré pattern is highlighted by a deformed hexagon, (d) 1 ML of silicon showing elliptical protrusions forming the Moiré pattern (U = -1V, I= 1 nA). The image shows the structure of the silicon layer in more detail. The Au [$\bar{1}$10] direction and the rectangular unit cell of 0.73 nm x 0.92 nm, rotated by 19° are also shown. All the directions shown on the figures correspond to Au directions.

In conclusion, the silicon 2D sheet synthesized on Au(111) unambiguously presents a Dirac cone at the high symmetry points of the BZ as measured by HR-ARPES. Moreover, these measurements show that the apex of the Dirac cone is 0.5 eV below the Fermi level. A single component of the Si 2p core level indicates a single silicon environment. Additionally, the XPS measurements rule out any alloying. We also find that the "herringbone" reconstruction of Au(111) remains after silicon deposition. Finally, these observations point to the formation of a silicene sheet weakly interacting with the Au substrate. The discovery of a silicene sheet on Au(111) may open the way for more integration in electronic devices[21].

**Methods:**

We used a commercial Au(111) crystal with 99.999% purity. The experiments were performed in ultra-high vacuum apparatus equipped with the tools for surface preparation and characterization: an ion gun for surface cleaning, Low Energy Electron Diffraction (LEED), low temperature Scanning Tunneling Microscopy (LT-STM, 5K) for the surface characterization at the atomic scale, and Auger Electron Spectroscopy (AES) for chemical surface analysis and silicon rate calibration. 1 ML of silicon corresponds to an attenuation of the gold Auger peak of about 40%.

The Au(111) sample was cleaned by several cycles of sputtering (600 eV Ar$^+$ ions, P ~ 10$^{-5}$ mbar ) followed by annealing at 450°C until a sharp (22 x √3) LEED pattern was obtained, signaling a clean Au(111) surface. Silicon was evaporated by direct current heating of a Si wafer onto the Au(111) surface held at 260°C. The crystal was transferred to the STM where all the images were acquired at 78 K. The STM experiments were performed at the ISMO-CNRS Institute. The photoemission measurements were performed at two Synchrotron Radiation Institutes; on the TEMPO beam-line at SOLEIL and on the APE beam-line at ELETTRA. The same Au(111) crystal was used for all the experiments.

**\*Corresponding authors:**

azzedine.bendounan@synchrotron-soleil.fr

hamid.oughaddou@u-psud.fr

**Author contributions:**

H.O. conceived and conducted the research project. S.S., H.E., A.B., P.K.D, and H.O. performed the experiments. H.E, A.B., P.K.D., A.K., A.P.S., F.S., I.V., A.J.M, G.D. and H.O analyzed the results and contributed to the scientific discussions and manuscript preparation.

**Competing financial interests:**

The authors declare no competing financial interests.